\documentclass[aps,prd,twocolumn,groupedaddress]{revtex4-1}
\usepackage{amsfonts}
\usepackage{mathrsfs}
\usepackage{amsmath}
\usepackage{xcolor}
\usepackage{amssymb}
\usepackage{bm}
\usepackage{extarrows}
\usepackage{slashed}
\usepackage{graphicx}

\usepackage[linktocpage=true,bookmarks=true,bookmarksnumbered=true,breaklinks=true,pdfpagemode=Fullscreen,pdfstartview=FitBH]{hyperref}

\newcommand{\n}{\nonumber}

\def\bge{\begin{equation}}
\def\ede{\end{equation}}
\def\bga{\begin{aligned}}
\def\eda{\end{aligned}}
\newcommand{\beq}{\begin{equation}}
\newcommand{\eeq}{\end{equation}}
\newcommand{\bq}{\begin{equation}}
\newcommand{\eq}{\end{equation}}
\newcommand{\ba}{\begin{array}}
\newcommand{\ea}{\end{array}}
\newcommand{\beqa}{\begin{eqnarray}}
\newcommand{\eeqa}{\end{eqnarray}}
\newcommand{\beqs}{\begin{subequations}}
\newcommand{\eeqs}{\end{subequations}}

\def\nn{\nonumber}

\def\dis{\displaystyle}

\def\({\left(}
\def\){\right)}

\def\End{\end{document}}

\def\hf{\frac{1}{2}}

\def\to{\rightarrow}

\def\End{\end{document}}

\begin{document}

\title{Connecting Leptonic Unitarity Triangle to Neutrino Oscillation}

\author{{\sc Hong-Jian He}\,$^{a,b,c}$ \,and\, {\sc Xun-Jie Xu}\,$^a$}
\affiliation{$^a$Institute of Modern Physics and Center for High Energy Physics,
             Tsinghua University, Beijing 100084, China\\
             $^b$Center for High Energy Physics, Peking University, Beijing 100871, China\\
             $^c$Kavli Institute for Theoretical Physics China, CAS, Beijing 100190, China\\
             (\,hjhe@tsinghua.edu.cn and xunjie.xu@gmail.com\,)
             }

\begin{abstract}
Leptonic unitarity triangle (LUT) provides a geometric description of
CP violations in the lepton-neutrino sector and is directly measurable in principle.
In this work, we reveal that the angles in the LUT have definite physical meaning,
and demonstrate the exact connection of the LUT to neutrino oscillations.
{\it For the first time,}
we prove that these leptonic angles act as phase shifts in neutrino oscillations,
by shifting $\,\Delta m^{2}L/2E$\, to $\,\Delta m^{2}L/2E+\alpha$\,,\, where
\,$(L,E,\alpha)$\, denote the baseline length, neutrino energy and
corresponding angle of the LUT. Each LUT has three independent parameters
and contains only partial information of the PMNS matrix. We demonstrate that
the partial information in each LUT can describe the corresponding neutrino oscillation.
Hence, for the first time, we uncover that
any given kind of neutrino oscillations contain at most three (rather than four)
independent degrees of freedom from the PMNS matrix.
This may provide a cleaner way for fitting the corresponding oscillation data.
\\[1.5mm]
\hfill
Phys.\ Rev.\ D (in Press) [arXiv:1311.4496]
%\\[2mm]
%\vspace*{5mm}
\end{abstract}

% 14.60.Pq Neutrino mass and mixing (see also 12.15.Ff Quark and lepton masses and mixing)
% 11.30.Er Charge conjugation, parity, time reversal, and other discrete symmetries
% 12.15.Ff Quark and lepton masses and mixing (see also 14.60.Pq Neutrino mass and mixing)
% 12.15.Hh Determination of Kobayashi¨CMaskawa matrix elements
% 12.15.Hh Determination of Cabibbo-Kobayashi & Maskawa (CKM) matrix elements
% 14.60.Lm Ordinary neutrinos
%\vspace*{5mm}

\maketitle

%%%%%%%%%---------Section 1-----------%%%%%%%%%%%%%

%\vspace*{3mm}
\noindent
{\bf 1.~Introduction}
\vspace*{3mm}

Nature has exhibited rich flavor structures
in both quark and lepton-neutrino sectors,
providing the sources of measurable CP violations which could be the origin
of observed matter-antimatter asymmetry in the Universe.
For quark sector, diagonalizing quark mass matrices leads to
Cabibbo-Kobayashi-Maskawa (CKM) mixing matrix \cite{CKM,CKM1} in charged currents.
The unitarity of CKM matrix generates six unitarity triangles (UT) \cite{PDG1}.
The angles of each triangle have clear physical meaning, and their nonzero values
directly signal the CP violation. For instance,
the most commonly used $\,d-b\,$ triangle is given by the relation,
\beqa
V_{ud}^{*}V_{ub}^{}+V_{cd}^{*}V_{cb}^{}+V_{td}^{*}V_{tb}^{} \,=\, 0 \,.
\label{eq:0529}
\eeqa
Its three angles \,$(\alpha,\, \beta,\, \gamma)$\,
can be directly measured in CP-violation experiments such as B meson decays.

In parallel to quark sector, the lepton-neutrino sector has
Pontecorvo-Maki-Nakagawa-Sakata (PMNS) mixing matrix \cite{Pon}
in charged currents. The neutrino oscillations are crucial for
testing the PMNS matrix, including its Dirac CP angle.
Measuring the CP asymmetry of neutrino oscillations,
$\,P[\nu_\ell^{}\to\nu_{\ell'}^{}] - P[\bar{\nu}_\ell^{}\to\bar{\nu}_{\ell'}^{}]\,$
($\ell\neq \ell'$),\,
is a direct probe of Dirac CP violation \cite{CP-AS,PDG},
and poses a major challenge to particle physics today.
An alternative and complementary method is to measure
the leptonic unitarity triangles (LUT) from neutrino oscillations.

Hence, our natural question is: in the leptonic sector,
{\it what is the physical meaning of those angles in the LUT and
how do they exactly connect to neutrino oscillations?}

In this work, we reveal that the angles of the LUT have definite physical meaning,
and derive the exact connection of LUT to neutrino oscillations.
We note that the LUT has only three independent parameters and does not
contain the full information of PMNS matrix;
but we will demonstrate that the three parameters of each LUT
are enough to describe the corresponding neutrino oscillations.  Especially,
for the first time, we will prove that {\it the angles of the LUT act as the phase-shifts
in the corresponding neutrino oscillation probabilities.}
Thus, for long baseline oscillation experiments with enough precision
to measure the distortion of energy spectrum, the angles in the LUT
may be directly extracted from the shift of the maximal
appearance point in the spectrum.
We also note that some other nice features of the LUT and their tests were studied
in the recent literature\,\cite{LUTx}.

%%%%%%%%%%%%%%%%%%%%%%%%%%%%%%%%%%%%%%%%%%%%%%%%%%%%%%%%%%%%%%%%%%%%%%%%%%%%%%%%%%%%%%%%%%%
%%%%%%%%%---------Section 2-----------%%%%%%%%%%%%%

\vspace*{4mm}
\noindent
{\bf 2.~Connecting LUT to Neutrino Oscillation}
\vspace*{3mm}

\begin{figure}
\centering
\includegraphics[scale=1]{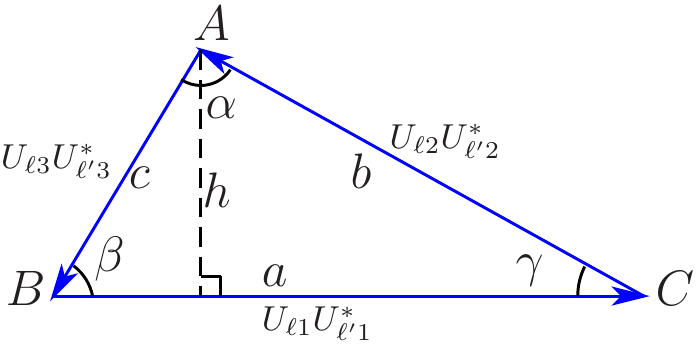}
\vspace*{-1mm}
\caption{The leptonic unitarity triangle (LUT), where $\,\ell\neq\ell'$,\,
\,$(a,\,b,\,c)$\, denote lengths of the three sides,
\,$(\alpha,\,\beta,\,\gamma)$\, denote the three angles,
and $\,h\,$ denotes the height.}
\vspace*{-1mm}
\label{fig:1}
\end{figure}

%\vspace*{2mm}
%\noindent
%{\it 2.~Connecting LUT to Neutrino Oscillation.} ---
Neutrinos are normally produced and detected in their flavor eigenstates
$\,|\nu_{\ell}^{}\rangle\,$ with $\,\ell = e,\mu,\tau\,$,\, which are mixtures of their
mass-eigenstates $\,|\nu_{j}^{}\rangle\,$  with $\,j=1,2,3$\,.\,
The flavor eigenstates $\,|\nu_{\ell}^{}\rangle\,$
and mass-eigenstates $\,|\nu_{j}^{}\rangle\,$ are connected
by the PMNS matrix $U$ \cite{Pon},\,
%
%$\,
\beqa
\vspace*{-4mm}
|\nu_{\ell}^{}\rangle \,=\, \sum_{j=1}^3 U_{\ell j}^{}|\nu_{j}^{}\rangle \,.\,
\vspace*{-3mm}
\eeqa
%$
%
Thus, a flavor state $\,|\nu_{\ell}^{}\rangle\,$ can oscillate into %may change into
$\,|\nu_{\ell'}^{}\rangle\,$  after flying a distance $\,L\,$.\,
The vacuum transition probability is given by \cite{CP-AS,PDG},
\begin{eqnarray}
\label{eq:U-lj}
\vspace*{-3mm}
P_{\ell\rightarrow \ell'}^{}
&\!=\!& \sum_{j=1}^{3}|U_{\ell'j}^{}U_{\ell j}^{}|^{2}
\\[-1mm]
&& +\,2\sum_{j<k}|U_{\ell'j}^{}U_{\ell j}^{}U_{\ell k}^{}U_{\ell'k}^{}|
   \cos (2\Delta_{jk}^{}\mp \phi_{\ell'\ell;jk}^{})\,,
\hspace*{8mm}
\n
\vspace*{-3mm}
\end{eqnarray}
where $\,\Delta_{jk}^{} \equiv {\,L\Delta m_{jk}^{2}\,}/(4E)\,$,\,
$\Delta m_{jk}^{2}$ is the mass-squared difference between
$\,|\nu_{j}^{}\rangle\,$ and $\,|\nu_{k}^{}\rangle\,$,\,
\,$E$\, denotes the neutrino energy, and
the $\,``\mp "\,$ signs correspond to $\,\nu_{\ell}^{}/\bar{\nu}_{\ell}^{}\,$
oscillations.
The phase angle $\,\phi_{\ell'\ell;jk}^{}\,$ is defined as \cite{PDG1,CP-AS},
\beqa
\phi_{\ell'\ell;jk}^{} ~\equiv~
\arg\(U_{\ell'j}^{}U_{\ell j}^{*}U_{\ell k}^{}U_{\ell'k}^{*}\) .
\label{eq:phi-def}
\eeqa
Thus we have,
$\,\phi_{\ell'\ell;jk}^{} = -\phi_{\ell\ell';jk}^{} = -\phi_{\ell'\ell;kj}^{}$\,
and $\,\Delta_{jk}^{}=-\Delta_{kj}^{}$\,.\,
Eq.\,\eqref{eq:U-lj} is a precise oscillation formula without
approximations \cite{PDG1,CP-AS}.
It also holds for $\,\ell=\ell'\,$,\,
which gives the survival probability of $\,\nu_{\ell}^{}\to\nu_{\ell}^{}\,$
($\bar{\nu}_{\ell}^{}\to\bar{\nu}_{\ell}^{}$)
with phase angle $\,\phi_{\ell\ell;jk}^{}=0\,$.\,
This survival probability ($\ell=\ell'$)
depends only on three parameters
$\,(\tilde{a}_\ell^{},\,\tilde{b}_\ell^{},\,\tilde{c}_\ell^{})\equiv
   (|U_{\ell 1}^{}|^2,\,|U_{\ell 2}^{}|^2,\,|U_{\ell 3}^{}|^2)\,$,
which obey the unitarity constraint of the matrix $U$,\,
$\,\tilde{a}_\ell^{} + \tilde{b}_\ell^{} + \tilde{c}_\ell^{} = 1\,$.\,
Hence, under $\,\ell =\ell'$,\, Eq.\,\eqref{eq:U-lj} actually
contains only {\it two independent degrees of freedom} among all four
parameters in the PMNS matrix.
For instance, we can express the disappearance probability in terms of
$(\tilde{a}_\ell^{},\,\tilde{b}_\ell^{})$, apart from $\,\Delta_{jk}$,\,
\begin{equation}
\begin{array}{l}
P_{\textrm{dis}}^{}
\,=\, 1-P_{\ell\rightarrow\ell}^{}
= 2\!\underset{j<k}{\Sigma}|U_{\ell j}^{}|^{2}|U_{\ell k}^{}|^{2}[1-\cos(2\Delta_{jk}^{})]
\\
\!\! =
 4\tilde{a}_{\ell}^{}\tilde{b}^{}_{\ell}\sin^{2}\!\Delta_{12}\!
+4(1\!-\tilde{a}_{\ell}^{}\!-\tilde{b}_{\ell}^{})
  (\tilde{a}_{\ell}^{}\sin^{2}\!\Delta^{}_{31}\! + \tilde{b}^{}_{\ell}\sin^{2}\!\Delta_{23}^{}).
\end{array}
\label{eq:P-dis}
\end{equation}
This clearly shows that the disappearance oscillations do not directly measure the LUT parameters
(cf.\ Fig.\,\ref{fig:1}), especially the LUT angles for CP-violation.
They probe the PMNS parameters only via the two independent quantities among
$\,(\tilde{a}_\ell^{},\,\tilde{b}_\ell^{},\,\tilde{c}_\ell^{})\,$,\, as shown above.
Hence, we will focus the present work on the appearance oscillations
($\ell\neq\ell'$), which contain nontrivial phase shift
$\,\phi_{\ell'\ell;jk}^{}\neq 0\,$.\,
Our key finding is to quantitatively connect the appearance oscillations to LUT's.

We note that the oscillation formulas \eqref{eq:U-lj}-\eqref{eq:phi-def}
mainly depend on the absolute values
such as $\,|U_{\ell j}^{}|$,\,
and $\,\phi_{\ell'\ell;jk}^{}\,$  is the {\it only place} where
complex phases of $\,U_{\ell j}^{}\,$ enter and generate observable CP violation
in neutrino oscillations.
We stress that, in contrast to the Dirac CP-phase $\,\delta\,$
in the conventional PMNS matrix \cite{PDG}, the phase angle
$\,\phi_{\ell'\ell;jk}^{}\,$ has the advantage of being parametrization-independent.
Furthermore, $\,\phi_{\ell'\ell;jk}^{}\,$ explicitly appears as the phase-angle-shift
in Eq.\,\eqref{eq:U-lj}, and may be directly read out from
the shift of maximal transition point in the neutrino energy spectrum
once the measurements become precise enough.

Then, we wish to ask:
what is the physical meaning of the phase shift
$\,\phi_{\ell'\ell;jk}^{}\,$? and how is it connected to the LUT\,?\,
Strikingly, we find that {\it the phase shift $\,\phi_{\ell'\ell;jk}^{}\,$
in neutrino oscillations is just one of the exterior angles in the LUT.}
This will be proven as follows.

%

%\vspace*{1mm}

The unitarity conditions of the PMNS matrix,
$\,U^{\dag}U=UU^{\dag}=1\,$,\, will result in two sets of LUT's,
$\,\sum_j^{}U_{\ell j}^{}U_{\ell'j}^* = 0\,$ with $\,\ell\neq\ell'\,$
(row triangles or ``Dirac triangles")
and $\,\sum_\ell^{}U_{\ell j}^*U_{\ell j'}^{}=0\,$ with $\,j\neq j'\,$
(column triangles or ``Majorana triangles").
For studying the flavor neutrino oscillations, we consider the ``Dirac triangles",
\beqa
U_{\ell 1}^{}U_{\ell'1}^{*} + U_{\ell 2}^{}U_{\ell'2}^{*} + U_{\ell 3}^{}U_{\ell'3}^{*}
\,=\, 0 \,,  &~~~& (\ell\neq \ell')\,.~~~
\label{eq:LUT}
\eeqa
This forms a triangle in the complex plane, as shown in Fig.\,\ref{fig:1}.
Its three sides have lengths,
\begin{eqnarray}
(a,\,b,\,c) & \,\equiv\, &
(|U_{\ell 1}U_{\ell'1}|,\,|U_{\ell 2}U_{\ell'2}|,\,|U_{\ell 3}U_{\ell'3}|) \,.~~~
\label{eq:abc}
\end{eqnarray}
The three angles are expressed as
\beqa
\label{eq:alpha-beta-gamma}
\alpha & \,=\, &
\arg\(\!-\frac{\,U_{\ell 2}^{}U_{\ell'2}^{*}\,}{U_{\ell 3}^{}U_{\ell'3}^{*}}\),
\nn
\\
\beta & \,=\, &
\arg\(\!-\frac{\,U_{\ell 3}^{}U_{\ell'3}^{*}\,}{U_{\ell 1}^{}U_{\ell'1}^{*}}\),
\\
\gamma & \,=\, &
\arg\(\!-\frac{\,U_{\ell 1}^{}U_{\ell'1}^{*}\,}{U_{\ell 2}^{}U_{\ell'2}^{*}}\),
\nn
\eeqa
In order to make exact connections to the phase angle \eqref{eq:phi-def},
we compute a generic arc angle of Eq.\,\eqref{eq:alpha-beta-gamma},
\begin{eqnarray}
\arg\!\(\! -\frac{U_{\ell j}^{}U_{\ell'j}^{*}}{U_{\ell k}^{}U_{\ell'k}^{*}}\)
 & = & \arg(U_{\ell j}^{}U_{\ell'j}^{*})-\arg(-U_{\ell k}^{}U_{\ell'k}^{*})
\nonumber \\ %[1.5mm]
 & = & \arg(U_{\ell j}^{}U_{\ell'j}^{*})+\arg(-U_{\ell k}^{*}U_{\ell'k}^{})
\nonumber \\[1.5mm]
 & = & \arg(U_{\ell j}^{}U_{\ell'j}^{*})+\arg(U_{\ell k}^{*}U_{\ell'k}^{}) + \pi
\nonumber \\[1.5mm]
 & = & \arg(U_{\ell'j}^{*}U_{\ell j}^{}U_{\ell k}^{*}U_{\ell'k}^{}) + \pi
\nonumber \\[1.5mm]
 & = & \pi - \phi_{\ell'\ell ;jk}^{} \,,
\label{eq:LUT-phi-1}
\end{eqnarray}
where we have used the identities,
$\,\arg(z_{1}^{}/z_{2}^{})=\arg(z_{1}^{})-\arg(z_{2}^{})$, $-\arg(z)=\arg(z^{*})$\,,\,
$\,\arg(-z)=\arg(z)+\pi$\,,\,
$\,\arg(z_{1}^{}z_{2}^{})=\arg(z_{1}^{})+\arg(z_{2}^{})$\,.\,
Note that all these equalities hold modulo $\,2n\pi\,(n\in\mathbb{Z})$.\,
Hence, we conclude that $\,\phi_{\ell'\ell;jk}^{}\,$
just equals one of the exterior angles in the LUT,
\beqa
\alpha = \pi-\phi_{\ell'\ell;23}^{},\,~~
\beta =  \pi-\phi_{\ell'\ell;31}^{},\,~~
\gamma = \pi-\phi_{\ell'\ell;12}^{}.~~~~~~
\label{eq:LUTa-phi}
\eeqa

\begin{figure}
\centering
\includegraphics[scale=1]{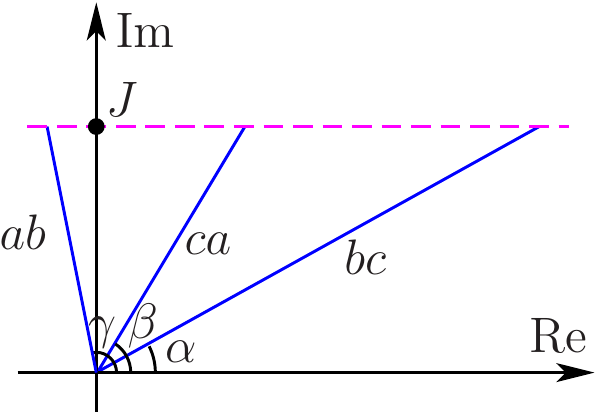}
\vspace*{-1mm}
\caption{Relation between $(\alpha,\,\beta,\,\gamma)$ and Jarlskog invariant \,$J\,$.\,
 The productions of any two sides of the triangle
 in the complex plane share the same imaginary part.}
\vspace*{-1mm}
\label{fig:2x}
\end{figure}

We further present a geometrical proof of the identities \eqref{eq:LUTa-phi}.
Let us first consider $\,\beta\,$ angle. From Fig.\ref{fig:1}, we have
\beqa
S_{\vartriangle}^{} = \hf a\,h\,,  ~~~~
\sin\beta \,=\, \frac{\,h\,}{c} \,=\, \frac{\,2S_{\vartriangle}\,}{ac} \,,~~~~
\label{eq:sinBeta}
\eeqa
where $\,S_{\vartriangle}^{}\,$ is the area of the triangle and
$\,h\,$ denotes the height. The Jarlskog invariant $J$ \cite{jar}
is the rephasing-invariant measure of CP violation and equals,
$\,J = \textrm{Im}(U_{\ell'j}^{}U_{\ell j}^{*}U_{\ell k}^{}U_{\ell'k}^{*})$\,,\,
where $\,\ell\neq \ell'\,$ and $\,j\neq k\,$.\,
Hence, we can derive the phase angle from \eqref{eq:phi-def},
\beqa
\sin\phi_{\ell'\ell;31}^{}
\,=\, \frac{J}{\,|U_{\ell'3}^{}U_{\ell 3}^{*}U_{\ell 1}^{}U_{\ell'1}^{*}|\,}
\,=\, \frac{J}{\,ac\,} \,. ~~~~~~
\label{eq:sinPhi-J}
\eeqa
Because each UT has its area equal half of Jarlskog invariant,
$\,S_{\vartriangle}^{}=J/2$\, \cite{foot-1},
the right-hand-sides of Eq.\,\eqref{eq:sinPhi-J} and
the second relation in Eq.\,\eqref{eq:sinBeta} are equal.
Hence, we arrive at,
\beqa
\sin\beta = \sin\phi_{\ell'\ell;31}^{} \,.\,
\label{eq:sinBeta-Phi}
\eeqa
Similarly, we deduce,
\beqa
\sin\alpha = \sin\phi_{\ell'\ell;23}^{}\,,
&~~~~~&
\sin\gamma = \sin\phi_{\ell'\ell;12}^{}\,.\,
\eeqa
These elegantly reprove our result (\ref{eq:LUTa-phi}) {\it in a geometrical way.}
It invokes the Jarlskog invariant, and also reveals a clear picture
for the relation between $\,(\alpha,\,\beta,\,\gamma)$\, and $\,J\,$.\,
We present this in Fig.\,\ref{fig:2x}, which demonstrates that the productions
of any two sides of the triangle in the complex plane share the same
imaginary part, i.e., the same height in Fig.\,\ref{fig:2x},
%The LUT geometry gives,
%
\beqa
bc\sin\alpha = ca\sin\beta = ab\sin\gamma = J \,.
\label{eq:sinJ}
\eeqa

Using Eqs.\,\eqref{eq:abc} and \eqref{eq:LUTa-phi}, we can express
the oscillation formula \eqref{eq:U-lj} fully in terms of the
geometrical parameters in the corresponding LUT,
\begin{eqnarray}
\label{eq:P-LUT}
\hspace*{-3mm}
P_{\ell\rightarrow \ell'}^{} &\,=~&
a^{2}+b^{2}+c^{2}-2ab\cos\!\( 2\Delta_{12}^{}\pm\gamma \)
\\[1.5mm]
\hspace*{-3mm}
 && -2bc\cos\!\( 2\Delta_{23}^{}\pm\alpha \)
    -2ca\cos\!\( 2\Delta_{31}^{}\pm\beta \) \!.~~~
\n
\end{eqnarray}
Note that $\,P_{\ell\rightarrow\ell'}(L=0)=0$\, holds
as expected, since the source neutrinos have no time to oscillate.
Thus, we can simplify the form of \eqref{eq:P-LUT} by subtracting
$\,P_{\ell\rightarrow\ell'}(L=0)$\,,
\beqa
P_{\ell\rightarrow \ell'}^{}
&=& 4ab\sin(\Delta_{12}\pm\gamma)\sin\Delta_{12}
\n\\[1mm]
\label{eq:P-LUT-f}
&& +4bc\sin(\Delta_{23}\pm\alpha)\sin\Delta_{23}
\\[1mm]
&& +4ac\sin(\Delta_{31}\pm\beta)\sin\Delta_{31} \,.
\n
\eeqa

Eqs.\,\eqref{eq:P-LUT}-\eqref{eq:P-LUT-f} demonstrate
the quantitative connection between $\,(\alpha,\,\beta,\,\gamma)$\,
of the LUT and the oscillation probabilities.
Hence, we have explicitly proven that {\it the physical meanings of
\,$(\alpha,\,\beta,\,\gamma)$\,
are just the phase shifts in the neutrino oscillations.}
It is striking to see that a flavor-changing oscillation ($\ell\neq \ell'$)
is {\it fully determined by the geometrical parameters of the LUT,}
for the given $\,\Delta m_{jk}^2$\, and
experimental setup $(E,\,L)$ \cite{foot-2}.

This has an important implication. Apart from two possible
Majorana phases, the PMNS matrix has \emph{four independent parameters}
(3 mixing angles and 1 Dirac CP angle), which
would all appear in the standard oscillation formula \eqref{eq:U-lj}.
But, a LUT has only \emph{three independent geometrical parameters}
and thus only contains partial information in the PMNS matrix.
Impressively, we have proven that this partial information of the PMNS matrix,
as contained in a given LUT \eqref{eq:LUT}, is enough to
determine the corresponding oscillation probability,
for the inputs $\Delta m_{jk}^2$ and \,$(E,\,L)$.\,

This feature is important for fitting an oscillation experiment
when higher experimental precision is reached such that all four parameters
of the PMNS matrix have observable effects.
In this case, we may suggest a 3-parameter fit based on each given LUT,
rather than the conventional 4-parameter fit in terms of
$\,(\theta_{12}^{},\,\theta_{13}^{},\,\theta_{23}^{},\,\delta)$\,
(which contains a redundant degree of freedom that cannot be determined
independently in a given kind of appearance experiments).
This has two advantages:
(i) the simplicity of \eqref{eq:P-LUT-f}
in terms of the geometric parameters of LUT;
(ii) the extra redundant degree of freedom in the conventional 4-parameter fit
of the PMNS matrix is automatically removed for a given kind of oscillation
experiments ($\ell\neq \ell'$).

Finally, since combining two different LUT's will provide the full information of the
PMNS matrix \cite{foot-3},
making two kinds of oscillation experiments can fit the two corresponding LUT's,
and thus give a full reconstruction of the PMNS matrix.

%%%%%%%%%%%%%%%%%%%%%%%%%%%%%%%%%%%%%%%%%%%%%%%%%%%%%%%%%%%%%%%%%%%%%%%%%%%%%%%%%%%%%%%%%%%
%%%%%%%%%---------Section 3-----------%%%%%%%%%%%%%

\vspace*{4mm}
\noindent
{\bf 3.~Probing the LUT via Neutrino Oscillations}
\vspace*{3mm}

In section, we further study how to test the LUT parameters via neutrino oscillations.
To determine a LUT, we can choose two sides plus one angle, say
$\,(a,\,b,\,\gamma )\,$,\, as the three independent geometrical parameters.
Then, all other parameters in this LUT can be expressed
in terms of $\,(a,\,b,\,\gamma )\,$,\,
\beq
\ba{ccl}
c^{2} &\,=\,& \dis a^{2}+b^{2}-2ab\cos\gamma \,,
\\[1.5mm]
\tan\alpha &\,=\,& \dis \frac{a\sin\gamma}{b-a\cos\gamma} \,,
\label{eq:c-beta-alpha}
\\[3mm]
\tan\beta &\,=\,& \dis \frac{b\sin\gamma}{a-b\cos\gamma} \,.
\ea
\eeq
Hence, we can reexpress \eqref{eq:P-LUT} or \eqref{eq:P-LUT-f} fully in terms of
$\,(a,\,b,\,\gamma )\,$,\,  although this makes the formula a bit lengthy.
But, for the realistic case of oscillation experiments, the formula
may be much simplified in terms of $\,(a,\,b,\,\gamma )\,$.\,
From the current oscillation data \cite{fit2013,Valle},
\beqa
\Delta m^{2} &\,\equiv\,&
|\Delta m_{13}^{2}| \,\simeq\, 2.4\times 10^{-3}\,\textrm{eV}{}^{2} \,,
\n \label{eq:0606}
\\[-3mm]
\\
\delta m^{2} &\,\equiv\,&
|\Delta m_{12}^{2}| \,\simeq\, 7.5\times 10^{-5}\,\textrm{eV}{}^{2} \,.
\n \label{eq:0606-1}
\eeqa
and considering the case of $\,E/L\sim\delta m^{2}$\,,\,  we find,
$\,\Delta_{12}^{} = O(1)\,$ and
$\,|\Delta_{23}|,|\Delta_{31}|\gg 1\,$.\,
Thus, the last two terms in \eqref{eq:P-LUT} will be averaged
out due to integration over the neutrino production region and
the energy resolution function, etc \cite{PDG1}. Hence, we deduce,
\beqa
P_{\ell\rightarrow \ell'}^{} &\,\simeq\,&
a^{2}+b^{2}+c^{2}-2ab\cos (2\Delta_{12}^{}\!\pm\gamma)
\n\\[2mm]
&\,=\,&
2(a^{2}+b^{2})-4ab\cos(\Delta_{12}^{}\!\pm\gamma)\cos\Delta_{12}^{}\,,
~~~~~~~~
\label{eq:P-abga}
\eeqa
which depends on $\,(a,\,b,\,\gamma )\,$ only.
In the above, the ``$\pm$" signs correspond to neutrino/antineutrino oscillations,
and we have used the first relation of \eqref{eq:c-beta-alpha} for deriving
the second equality of \eqref{eq:P-abga}.

\begin{figure}[t]
\vspace*{2mm}
\begin{center}
\includegraphics[width=8.5cm,height=6.0cm]{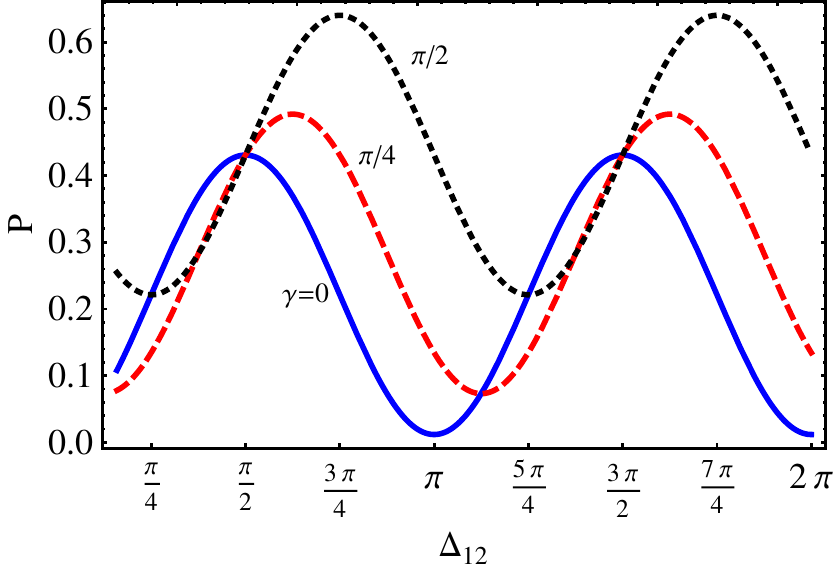}
\vspace*{-3mm}
\caption{Phase-shift effects of $\,\gamma\,$ on neutrino oscillation
probability $\,P[\bar{\nu}_\ell^{}\to\bar{\nu}_{\ell'}^{}]$\,.\,
For illustration, we plot three curves for
$\,\gamma=0\,$ (blue solid), $\,\frac{\pi}{4}\,$ (red dashed), and
$\,\frac{\pi}{2}$\, (black dotted).}
\label{fig:2}
\end{center}
\vspace*{-6mm}
\end{figure}

In Eq.\,\eqref{eq:P-abga},
if $\,\gamma =0$\,,\, the $\,\nu_{\ell'}^{}\,$ maximal appearance point
is  $\,\Delta_{12}^{}=\frac{\pi}{2}$\,.\,
For a nonzero $\,\gamma$\,,\, the maximal appearance point is shifted to
\begin{equation}
\Delta_{12}^{\star}
\,=\, \frac{\pi}{2}\mp\frac{\gamma}{2} \,,
\label{eq:0605-4}
\end{equation}
and its corresponding appearance probability is
\beqa
P_{\ell\rightarrow \ell'}^{\max} \,\simeq\,
2(a^{2}+b^{2}) + 4ab\sin^2\frac{\gamma}{2} \,.
\label{eq:P-max}
\eeqa
This phase-shift effect is shown in Fig.\,\ref{fig:2}
for $\,\bar{\nu}_\ell^{}\to\bar{\nu}_{\ell'}^{}$\, oscillations,
where we use \eqref{eq:P-abga} with sample inputs
$\,(a,\,b)=(0.29,\,0.36)\,$.\, The three curves in Fig.\,\ref{fig:2} correspond to
$\,\gamma = \(0,\,\frac{\pi}{4},\,\frac{\pi}{2}\)$,\,
and have their first maximal appearance points located at
$\,\Delta_{12}^{\star}=\(\frac{\pi}{2},\,\frac{5\pi}{8},\,\frac{3\pi}{4}\)$,\,
in accord with \eqref{eq:0605-4}.
Fig.\,\ref{fig:2} also shows that the curves move upward with the increase
of $\,\gamma\,$.\, This can be understood from the maximal appearance probability \eqref{eq:P-max}
which monotonously rises with the increase of $\,\gamma\in (0,\,\pi)\,$.\,
We have made similar analyses for choosing other input parameters of the LUT,
such as $\,(b,\,c,\,\alpha )\,$ and $\,(a,\,c,\,\beta )\,$.

In addition, using Eq.\,\eqref{eq:P-abga}, we can
derive the probability difference between the neutrino and antineutrino oscillations,
\beqa
\label{eq:CPas}
 P_{\ell\rightarrow \ell'}^{} \!-\!
 P_{\bar{\ell}\rightarrow \bar{\ell}'}^{}
& \,\simeq\, & 4ab \sin\gamma \sin (2\Delta_{12}^{})
\\[2mm]
& \,=\, &
8S_{\vartriangle}^{}\sin (2\Delta_{12}^{})
\,=\, 4J \sin (2\Delta_{12}^{}) \,.~~~~~~
\nonumber
\eeqa
The CP asymmetry (\ref{eq:CPas}) provides the net measure of CP violation
in terms of the area of the LUT, $\,S_{\vartriangle}^{}=J/2$\,,\, as expected.

From Eqs.\,\eqref{eq:P-abga}-\eqref{eq:P-max},
we see that the angle $\,\gamma\,$ plays the physical role of phase shift in neutrino
oscillations with $\,E/L \sim \delta m^2\,$.\,
In principle, we can change either $\,L\,$ or $\,E\,$
to detect how the maximal appearance point is shifted,
and thus directly measure $\,\gamma$\,.\,
In practice, it is much easier to vary $\,E\,$
since moving around a large detector would be hard.

Actually, a more realistic method is to measure the distortion of
neutrino energy spectrum. For instance, we may produce many muon neutrinos
$\nu_{\mu}^{}$ with different energies
which can be measured or are already known.
Then, at the far detector with $\,L\sim E/\delta m^{2}\,$,\,
we will measure the $\,\nu_{e}^{}\,$ appearance with a different energy spectrum.
Thus, we may use (\ref{eq:P-abga}) to fit
the distortion of the spectrum and infer the values of
$\,(a,\,b,\,\gamma )\,$ in the $e-\mu$ LUT.
The current $\,\nu_{\mu}^{}\rightarrow\nu_{e}^{}\,$ experiments
cannot reach such a small $\,E/L\,\sim \delta m^{2}\,$.\,
For instance, MINOS experiment \cite{minos} has $\,E/L\,$ about
$\,3\textrm{GeV}/735\textrm{km}\simeq 8\times 10^{-4}\,\textrm{eV}^{2}$\,
which is insensitive to the oscillations via $\,\Delta_{12}\,$.\,
The situation of NO$\nu$A experiment \cite{NOvA} is similar, which has
$\,E/L\simeq 2\,\text{GeV}/810\,\text{km}\simeq 5\times 10^{-4}\,\textrm{eV}^{2}$.\,
The Super Beam Project \cite{EURNu} creates 300\,MeV muon
neutrinos and has a 130\,km baseline, with
$\,E/L\simeq 4.5\times 10^{-4}\,\textrm{eV}^{2}\,$
also at the same order as MINOS experiment.
The future Neutrino Factory \cite{Nufac}
will have baseline $\,L = 2000-7500$\,km and neutrino energy $\,E = O(1-10)$GeV,
which is possible to realize $\,E/L\sim\delta m^{2}\,$.

The $\,\nu_{\mu}^{}\rightarrow\nu_{e}^{}\,$ oscillation experiments measure
the appearance probability $\,P_{\mu\rightarrow e}^{}(E)$\, in Eq.\,(\ref{eq:P-abga})
as a function of neutrino energy $\,E\,$ in a long baseline $\,L \sim E/\delta m^{2}\,$.\,
To inspect the sensitivity of $\,P_{\mu\rightarrow e}^{}(E)$\, to the angle $\,\gamma\,$,\,
we first evaluate the ranges of $\,(\gamma,\,a,\,b)$\,
in the $e-\mu$ LUT from the present oscillation data.
The new global fit of the PMNS matrix gives \cite{fit2013},
\begin{equation}
\begin{array}{ccl}
%\hspace*{-4mm}
s_{12}^{2} &=& (3.08 \!\pm\! 0.17) \!\times\! 10^{-1}\!,~
\\[1.5mm]
s_{23}^{2} &=& (4.25 \!\pm\! 0.28) \!\times\! 10^{-1}\!,
\\[1.5mm]
%\hspace*{-4mm}
s_{13}^{2} &=& (2.34 \!\pm\! 0.20) \!\times\! 10^{-2},~
\\[1.5mm]
\delta &=& (1.39 \pm 0.30)\pi \,,
%\vspace*{-4mm}
\end{array}
\label{eq:0613}
%\vspace*{3mm}
\end{equation}
where $s_{ij}^{2}\equiv\sin^{2}\theta_{ij}$\, and
$\,\pm 1\sigma\,$ errors are included.
The PMNS matrix can be expressed as, $\,U=U_0^{}U'$, with
{\footnotesize
\beq
%\vspace*{-4mm}
\hspace*{-1.5mm}
U_0^{} \!=\!\! \left\lgroup\!\!
\begin{array}{ccl}
c_{31}c_{12} & c_{31}s_{12} & s_{31}e^{-i\delta}
\\
-s_{12}c_{23} \!-\! c_{12}s_{23}s_{31}e^{i\delta}
& c_{12}c_{23} \!-\! s_{12}s_{23}s_{31}e^{i\delta} & s_{23}c_{31}
\\
s_{12}s_{23} \!-\! c_{12}c_{23}s_{31}e^{i\delta}
& -c_{12}s_{23} \!-\! s_{12}c_{23}s_{31}e^{i\delta} & c_{23}c_{31}
\end{array} \!\! \right\rgroup \!\! .
\label{eq:PMNS}
\eeq
}
\hspace*{-2.5mm}
The Majorana phase matrix
$\,U'=\textrm{diag}(1,\,e^{i\varphi_2^{}},\,e^{i\varphi_3^{}})\,$
does not affect the ``Dirac triangles" \eqref{eq:LUT} and is irrelevant to
the oscillation analyses.
Using Eqs.\,\eqref{eq:abc}-\eqref{eq:alpha-beta-gamma} and the mixing matrix
\eqref{eq:PMNS}, we can reconstruct the LUT parameters $\,(\gamma,\,a,\,b)$\,
from the neutrino data \eqref{eq:0613}.

We present the probability distributions of $\,(\gamma,\,a,\,b)$\,
in Fig.\,\ref{fig:3}(a)-(c).
We have simulated 30000 samples in each plot and normalized the total area of each
histogram as unit.  We find that $\,\gamma\,$ falls into a narrow range,
\beqa
-20^{\circ}\lesssim \gamma \lesssim 20^{\circ}\,,\,
\eeqa
with a most probable value $\,\gamma\simeq 15.5^\circ\,$.\,
Fig.\,\ref{fig:3} further constrains,
\beqa
0.25\lesssim a \lesssim 0.45\,,\,
&~~~&
0.29\lesssim b \lesssim 0.42\,,\,
\eeqa
with the most probable values $\,(a,\,b) \simeq  (0.29,\,0.36)\,$.\,

%\vspace*{1mm}

%
\begin{widetext}
\begin{center}
\begin{figure}[t]
%\vspace*{-1mm}
\includegraphics[width=17.5cm,height=5.2cm]{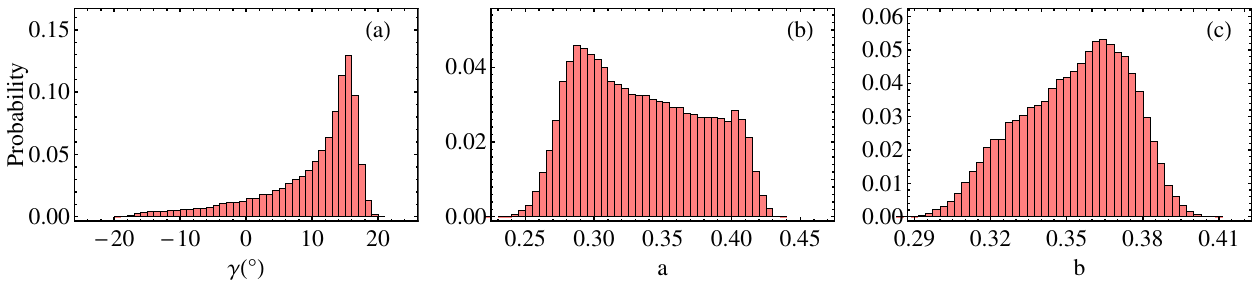}
\vspace*{-1mm}
\caption{Probability distributions of the geometric parameters in the $e-\mu$ LUT,
$\,\gamma\,$ [plot-(a)], $\,a\,$ [plot-(b)], and $\,b\,$ [plot-(c)],
based on the current neutrino global fit \cite{fit2013}.
We have simulated 30000 samples in each plot. }
\label{fig:3}
\vspace*{-8mm}
\end{figure}
\end{center}
\end{widetext}

\begin{figure}[b]
\begin{center}
\vspace*{1.5mm}
\includegraphics[width=8.5cm,height=6.0cm]{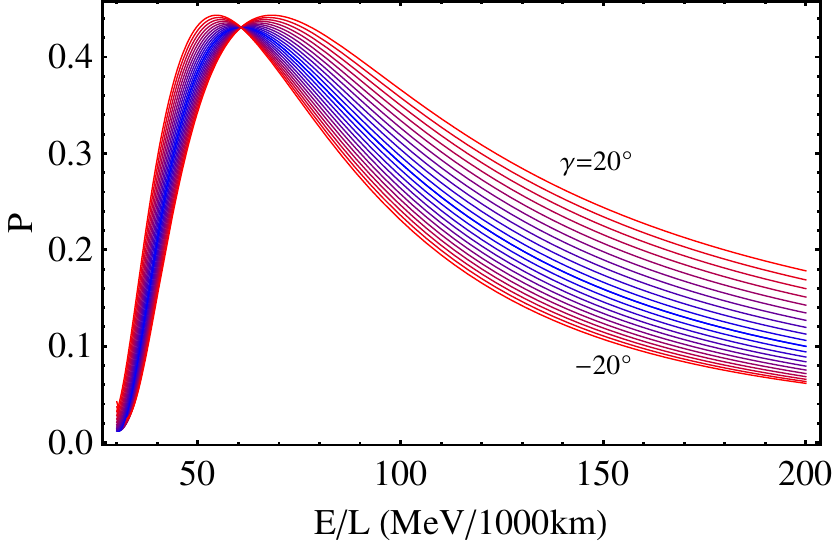}
\vspace*{-1mm}
\caption{The $\nu_{\mu}^{}\rightarrow\nu_{e}^{}$ oscillation probability versus
$\,E/L\,$,\,
for illustration of long baseline oscillations with $\,E/L\sim \delta m^{2}\,$.\,
Different curves (from bottom to top) correspond to varying $\gamma$
within $\,[-20^{\circ},\,20^{\circ}]$\, and with steps of \,$2^{\circ}$.}
\label{fig:4}
\end{center}
\vspace*{-5mm}
\end{figure}

In Fig.\,\ref{fig:4}, we plot the oscillation probability
$\,P_{\mu\rightarrow e}^{}(E)\,$ as a function of $\,E/L\,$,\,
based upon \eqref{eq:P-abga},
where we vary $\,\gamma\,$ values in the range $\,[-20^{\circ},\, 20^{\circ}]\,$
with steps by $\,2^{\circ}$.\,
We also set the sample inputs $\,(a,\,b) = (0.29,\,0.36)\,$
from their most probable values in Fig.\,\ref{fig:3}.
Fig.\,\ref{fig:4} shows that with $\,\gamma\,$ changing from
$\,-20^{\circ}\,$ to $\,20^{\circ}\,$,\, the maximum point of
$\,P_{\mu\rightarrow e}^{}\,$ shifts from left to right, and
its tail on the right-hand-side lifts up.
Fig.\,\ref{fig:4} clearly illustrates how $\,\gamma\,$
plays the physical role of phase shift.

\vspace*{1mm}

We note that observing the phase-shift effects is based on a premise that $\,L/E\,$
is \emph{variable} in the experiments. Hence, to probe the phase-shift effects requires
experiments to reach a relatively high resolution on the neutrino energy and the energy spectrum,
so the shape of the distribution in Fig.\,\ref{fig:4} (by varying energy $\,E$) can be measured.
Thus, the LUT parameters \,$(\gamma,\,a,\,b)$\, can be inferred from
fitting the measured energy spectrum.

\vspace*{1mm}

In passing, we also note that determining CP violation in an experiment with \emph{fixed} $L/E$
also involves the parameter degeneracy problem \cite{P-DG}, such as the $(\theta_{13},\,\delta)$ degeneracy,
implying that the oscillation probability for one pair of inputs $(\theta_{13},\,\delta)$ may equal that
for another pair $(\theta_{13}',\,\delta')$.\,
This problem is inherent in the three-neutrino oscillations and cannot be removed by simply enhancing the accuracy.
It may be resolved by varying $L/E$, {\it e.g.,} combining data from experiments
with different baselines and channels, or making use of the energy spectrum.

\vspace*{1mm}

Finally, we comment on the matter effects\,\cite{MSW}\cite{PDG1}.
To effectively measure the phase shift effects,
we should check the required size of \,$E/L$.\,
According to the above discussions, a relatively small $\,E/L \sim \delta m^{2}\,$\, is needed, which is beyond
the current experimental setup. For instance, the 735\,km baseline of MINOS\,\cite{minos} would need a neutrino
beam energy $\,E\sim 100$\,MeV for a sensitive probe.
In such case, the matter effect is only about $1/30$ of that involved in the current
MINOS setup (with $\,E \simeq 3$\,GeV), and thus negligible.
The case of NO$\nu$A \cite{NOvA} (with $\,E/L \simeq 2\,\textrm{GeV}/810\,\text{km}$)
is similar.
Besides, the Super Beam Project \cite{EURNu} has $\,E/L \simeq 300\,\textrm{MeV}/130\,\text{km}$,\,
whose neutrino energy is about a factor $1/10$ lower than the current MINOS setup,
so its matter effect will be insignificant.
This means that our formula (\ref{eq:P-abga}) would
give fine approximation for $\delta m^2$ dominated oscillations with  $\,L<1000$\,km.
Only for experiments with very long baselines (well above $1000$\,km) and high precision,
the matter effect would become sizable for probing the LUT's;
but this is fully beyond our current scope and
we will pursue such elaborated applications elsewhere.

%%%%%%%%%%%%%%%%%%%%%%%%%%%%%%%%%%%%%%%%%%%%%%%%%%%%%%%%%%%%%%%%%%%%%%%%%%%%%%%%%%%%%%%%%%%
%%%%%%%%%---------Section 4-----------%%%%%%%%%%%%%

\vspace*{5mm}
\noindent
{\bf 4.~Conclusions}
\vspace*{3mm}

Discovering leptonic CP violation has vital importance for neutrino physics,
as it may provide the origin of the observed matter-antimatter asymmetry
in the Universe \cite{BA-Rev}.
The leptonic unitarity triangle (LUT) provides a geometric description of
CP violations in the lepton-neutrino sector and is directly measurable.
Finding any nonzero angle of the LUT will be a direct proof of the
leptonic CP violation \cite{foot-1}.

\vspace*{1mm}

In this work, we revealed that the angles in the LUT have definite physical meaning,
and they act as the phase shifts of neutrino oscillations.
{\it For the first time,} we proved that the oscillation phases $\,\phi_{\ell'\ell;jk}^{}\,$
in the conventional formula \eqref{eq:U-lj} exactly
equal the corresponding exterior angles of the LUT, as in Eq.\,\eqref{eq:LUTa-phi}.
Our proof uncovers that a given kind of appearance oscillations
can be described by the corresponding LUT
with only three independent geometric parameters \cite{foot-2}.
This may provide a cleaner way for fitting the neutrino oscillation data,
since each kind of long baseline oscillation \eqref{eq:U-lj} is traditionally described
by four independent parameters in the PMNS matrix \eqref{eq:PMNS}.

\vspace*{1mm}

Without losing generality, we considered the $\,\nu_{\mu}^{}\rightarrow\nu_{e}^{}\,$ oscillations
with a long baseline $\,L\sim E/\delta m^{2}\,$,\, and studied
one of the LUT angles $\,\gamma\,$ for illustration.
We demonstrated that the oscillation formula takes a simple form (\ref{eq:P-abga}),
depending only on the three independent geometric parameters
$\,(\gamma,\,a,\,b)\,$ of the unitarity triangle.
We explicitly analyzed how the maximal appearance point of $\,\nu_{\mu}^{}\rightarrow\nu_{e}^{}\,$
oscillations gets shifted when $\,\gamma\,$ changes, as shown in Eq.\,(\ref{eq:0605-4})
and Figs.\,\ref{fig:2},\ref{fig:4}.
We will apply this LUT method to study concrete long baseline oscillation experiments elsewhere.

\vspace*{3mm}
\noindent
{\bf Acknowledgements.}
%\\[1.5mm]
%\hspace*{3mm}
We thank Samoil M.\ Bilenky, Eligio Lisi, Hitoshi Murayama, Werner Rodejohann, and
Zhi-zhong Xing for valuable discussions on testing leptonic unitarity triangles
after posting this paper to arXiv:1311.4496.
This work was supported by Chinese NSF (Nos.\ 11275101 and 11135003)
and National Basic Research Program (No.\ 2010CB833000).

%\vspace*{-2mm}


\begin{thebibliography}{References}
\bibitem{foot-1}
The relation $\,S_{\vartriangle}^{}=J/2$\, shows that the area of UT
is a (rephasing-invariant) direct measure of CP violation, and all six UT's
have equal area. Hence, an experimental finding of any nonzero angle in the UT
will be a direct proof of CP violation. In addition, finding the longest side
of the UT to be shorter than the sum of two other sides could also prove the CP violation.
The PMNS matrix contains two possible Majorana phases
in general, $\,U=U_0^{}U'\,$,\, where $\,U_0^{}\,$ consists of three mixing angles
and one Dirac CP phase as in Eq.\,\eqref{eq:PMNS}, and
$\,U'=\textrm{diag}(1,\,e^{i\varphi_2^{}},\,e^{i\varphi_3^{}})\,$
is the diagonal Majorana phase matrix. Note that the ``Dirac triangles" \eqref{eq:LUT}
are independent of the Majorana phase matrix $\,U'\,$,\, while the Majorana phases of $\,U'$
could only change the orientations of the ``Majorana triangles"
$\,\sum_\ell^{}U_{\ell j}^*U_{\ell j'}^{}=0\,$ ($\,j\neq j'\,$),
but leave their areas, sides and angles invariant.
\end{thebibliography}
\end{document}